# Core-Shell Bimetallic Nanoparticle Trimers for Efficient Light-to-Chemical Energy Conversion

Seunghoon Lee,[1,2] Heeyeon Hwang,[1] Wonseok Lee,[1] Dmitri Scherbarchov,[3] Younghyun Wy,[1] Johan Grand,[3,4] Baptiste Auguié,[3] Dae Han Wi,[1] Emiliano Cortés,[2,*] and Sang Woo Han[1,*]

[1] Center for Nanotectonics, Department of Chemistry and KI for the NanoCentury, KAIST, Daejeon 34141, Korea

[2] Nanoinstitute Munich, Faculty of Physics, Ludwig-Maximilians-Universität München, 80539 Munich, Germany

[3] The MacDiarmid Institute for Advanced Materials and Nanotechnology, School of Chemical and Physical Sciences, Victoria University of Wellington, P.O. Box 600, Wellington 6140, New Zealand

[4] Université de Paris, ITODYS, CNRS, F-75006, Paris, France

**ABSTRACT:** Incorporation of catalytically active materials into plasmonic metal nanostructures can efficiently merge the reactivity and energy harvesting abilities of both types of materials for visible light photocatalysis. Here we explore the influence of electromagnetic hotspots in the ability of plasmonic core-shell colloidal structures to induce chemical transformations. For this study, we developed a synthetic strategy for the fabrication of Au nanoparticle (NP) trimers in aqueous solution through fine controlled galvanic replacement between Ag nanoprisms and Au precursors. Core-shell Au@M NP trimers with catalytically active metals (M = Pd, Pt) were subsequently synthesized using Au NP trimers as templates. Our experimental and computational results highlight the synergy of geometry and composition in plasmonic catalysts for plasmon-driven chemical reactions.

**TOC GRAPHICS**

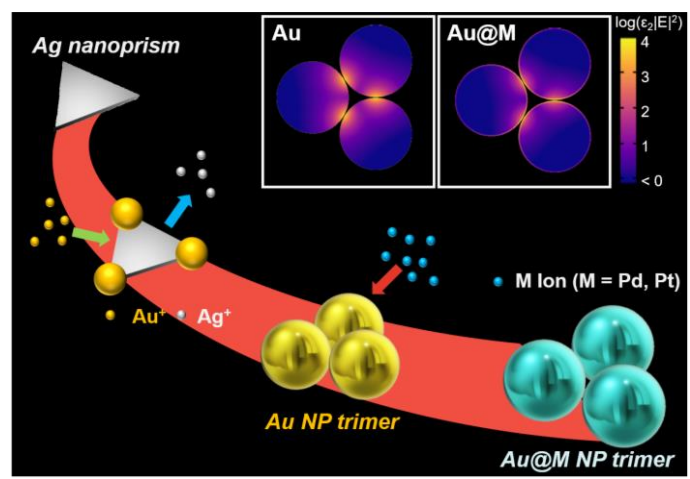

Plasmonic nanoparticles (NPs) interact strongly with light through the excitation of localized surface plasmon resonances (LSPRs); this effect has drawn great interest in various research fields such as bio/chemical sensing, photovoltaics, and thermal science, among others.[1-10] In particular, the utilization of plasmons for catalysis has raised attention in recent years as it may open unprecedented chemical transformation pathways and/or catalyze chemical reactions employing milder reaction conditions.[9–18] Plasmonic photocatalysts rely on new or up-to-now scarcely explored mechanisms for converting light into chemical energy. However, despite tremendous advances in recent years, we still lack a complete and unified scenario for the energy transfer mechanism across the plasmonic-molecular interface. Unraveling such a route could aid in the design of optimal photocatalysts.[18–23] For example, from the plasmonic viewpoint, it is well known that assembled nanostructures (i.e., nanostructures with hotspots) can highly enhance photon absorption, charge-carriers generation, electric-field (e-field) confinement, and temperature.[1,4,24–26] All these effects can highly improve the performance of plasmonic NPs functioning as catalysts.[10]

Among assembled nanostructures, plasmonic NP oligomers (NPOs), such as dimers, trimers, and tetramers, enable the systematic study of physicochemical phenomena with well-defined number of particles and geometrical arrangements.[27–37] There are numerous strategies for the formation of NPOs, such as DNA-mediated interactions, solvent/ligand interactions, and nanolithography.[1,2,27–37] Despite the various preparation routes and applications, NPOs have been rarely tested as photocatalysts in colloidal solution due to the lack of producing them with sufficient structural stability and controllable composition.[1,33,34] In this sense, for plasmonic catalysis, both structure (i.e. such as plasmonic oligomers with hotspots) and composition (i.e. hybrid structures) are central aspects to be considered.

In recent years, the incorporation of catalytically active metals (e.g., Pd, Pt, Rh, Ru) into plasmonic nanostructures — to form bimetallic nanostructures, such as alloys, antenna-reactor complexes, or core-shell NPs — has shown to be of potential interest for plasmonic photocatalytic applications. These systems successfully combine versatile catalytic functions — in the adsorption/desorption of molecules or lowering the activation barriers of the reaction — and efficient photon harvesting abilities.[38–47] Despite recent successful examples in the design and uses of some of these bimetallic nanostructures for visible light photocatalysis, their optimization as photocatalysts requires further understanding of the physical processes behind the light-into-chemical energy conversion in these systems.

Here, we systematically investigate the benefits of tuning structure and composition in bimetallic assembled NPs for plasmonic photocatalysis. We present an aqueous synthetic method for the preparation of Au, Au@Pd, and Au@Pt NP trimers with remarkable structural stability. Recently, we developed efficient assembly strategies to produce stable colloidal assembled nanostructures, such as Au NP clusters (NPCs), core-shell NPCs, and particle-in-a-frame nanostructures with well-defined geometrical parameters, based on the galvanic replacement of Ag nanoprisms with Au precursors.[7,8,46] These previous findings prompted us to explore the possibility of synthesizing Au and Au@M (M = Pd, Pt) NP trimers in aqueous solution. Distinctive from our prior works involving $Au^{1+}/I^{1-}$ complexes to generate Au NPCs,[7,8,46] we synthesized Au NP trimers by lowering the reaction temperature and by employing a weaker reducing agent, which maximizes a site-selective type of growth. The Au@Pd and Au@Pt NP trimers were successfully synthesized using Au NP trimers as seeds. Our experimental measurements and computational modeling showed an enhancement in the shell absorption for the hybrid trimers in comparison to their monomers, dimers, and monometallic counterparts. Finally, we monitored the plasmon-mediated reduction of 4-nitrobenzenethiol (4-NBT) to 4,4′-dimercapto azobenzene (DMAB), highlighting the importance of controllable structure and composition in driving plasmon-induced chemical reactions.

We start by describing the synthesis and growth mechanism of the colloidal Au NP trimers. To synthesize the Au NP trimers, we prepared an aqueous growth solution containing $HAuCl_4$, KI, and poly(vinyl pyrrolidone) (PVP, a capping/stabilizing agent) that was initially set at -2 °C in a low-temperature reactor. Afterward, hydroxylamine hydrochloride ($NH_2OH·HCl$) was added to this solution. In a typical synthesis, Ag nanoprisms with an average edge length of 35 nm, as synthesized by a previously-reported method,[48] were quickly injected into the growth solution and acted as seeds for the trimers formation. The mixture was left undisturbed in the low-temperature reactor for 60 min, after which the samples were exposed to room temperature. The color of the mixture changed from yellowish to purple [see the Supporting Information (SI) for further experimental details]. Transmission electron microscopy (TEM) and scanning electron microscopy (SEM) were used to characterize the products and confirmed the formation of the Au NP trimers (Figure 1a and b). The Au composition in the trimers — determined by energy-dispersive X-ray spectroscopy (EDS) — was found to be 99%. These reaction conditions showed a 68% yield for the Au NP trimers, which was estimated from the TEM images of the products of the reaction (Figures S1a and S2 in the SI). The average particle and gap sizes of the Au NP trimers were 58 ± 6 and 0.62 nm, respectively (Figure S1b and c in the SI). The preservation of the NP trimer structure in colloidal solution can be attributed to the van der Waals interaction between the constituent NPs, which is mediated by PVP.[7,8,46] The presence of residual PVP in the Au NP trimers and the removal of PVP from the Au NP trimers were identified by Fourier transform infrared (FTIR) spectroscopy measurements (Figure S3 in the SI). In fact, when the Au NP trimers were treated with an excessive amount of $NaBH_4$, which can act as a role of removing PVP from the surface of NPs,[49] the Au NP timers were destroyed and aggregated each other (Figure S3c-e in the SI),

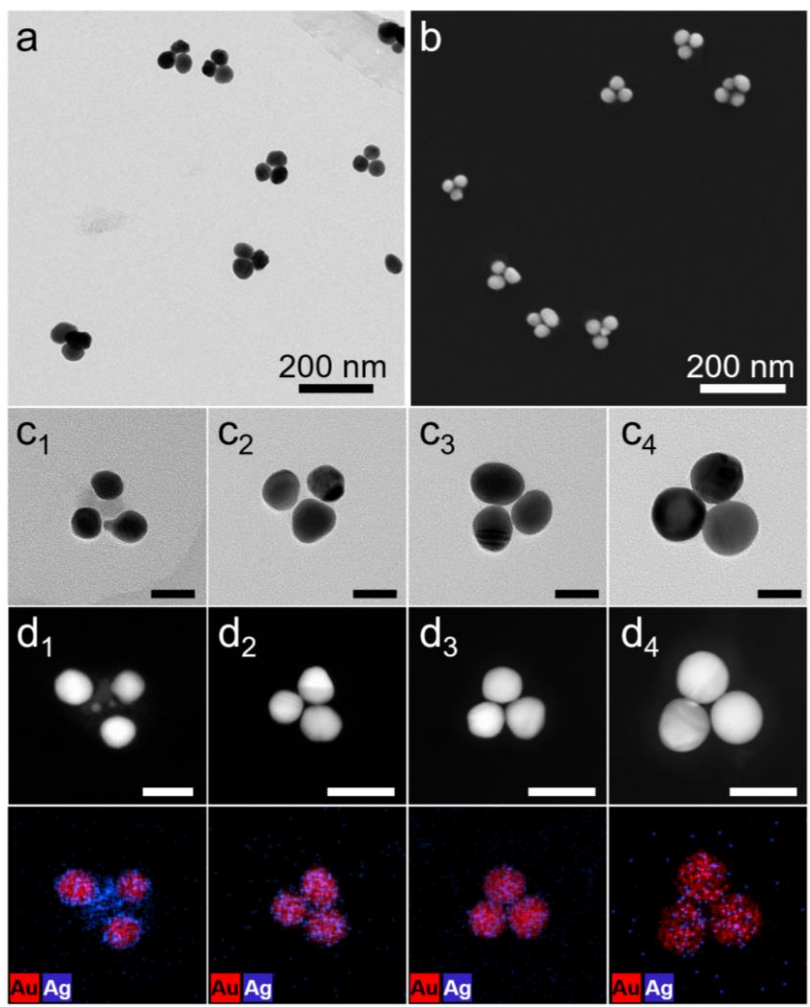

**Figure 1.** (a) TEM and (b) SEM images of Au NP trimers. (c) TEM images of samples collected at different reaction times during the formation of Au NP trimers: (c1) 0, (c2) 10, (c3) 30, (c4) 60 min. The scale bars indicate 30 nm. (d) HAADF-STEM images and corresponding EDS elemental mapping images of Au NP trimers collected at different reaction times: (d1) 0, (d2) 10, (d3) 30, (d4) 60 min. The scale bars indicate 40 nm.

In order to clarify the Au NP trimers formation mechanism, we performed a time-dependent TEM imaging study together with a high-angle annular dark-field scanning TEM (HAADF-STEM)-EDS elemental mapping (Figure 1c and d). From these images, we could deduce the growth mechanism, which we summarize as follows. In the initial growth stage — i.e., the sample was taken right after adding the Ag nanoprism seeds — we observed that the Au NPs start being formed at the corners of the nanoprisms (Figure 1c1 and d1). This can be due to the selective galvanic replacement at the corner areas of the Ag nanoprism seeds, which are highly energetic sites compared to other regions of the nanostructure.[50] As time goes on, the further growth of each Au corner NPs takes place by the reduction of the Au precursors with reducing agents (Figure 1c1-c4 and 1d1-d4). Unlike the previous strategies,[7,8,46] we conducted the galvanic replacement reactions at -2 °C using a weak reducing agent ($NH_2OH·HCl$) to decelerate the kinetics of the galvanic replacement reaction, which could maximize the site-selective growth of Au to produce Au NP trimers. Notably, the same reaction performed at elevated temperatures, such as 24, 60, and 85 °C, resulted in a strong decrease in the yield of Au NP trimers or even the synthesis of single NPs (Figure S4 in the SI). When the reaction was performed in the absence of $NH_2OH·HCl$ and/or KI, Au NP trimers could not be synthesized (Figure S5 in the SI). Based on these results, we show in Scheme 1 the proposed growth mechanism for the Au NP trimers.

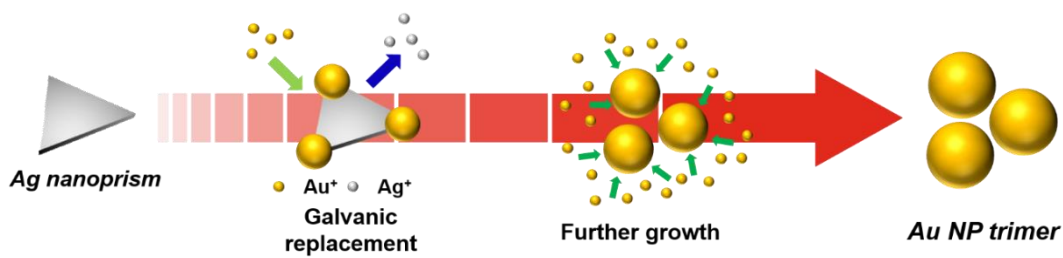

**Scheme 1.** Schematic illustration of the growth mechanism of Au NP trimers.

After the synthesis of the Au NP trimers, we coated them with a thin layer of catalytically active metals: Pd and Pt. The adsorption of molecules on these partially filled d-band metals generally reduces the activation barrier of chemical reactions and/or facilitates desorption of products of the reaction from the catalyst's surface. In this way, our hybrid structures aim to combine the light harvesting capabilities of plasmonic NPOs with catalytically active metals.[13,14] Note that thin catalytic layers would still allow the plasmon excitation of the cores.[44] The core-shell NP trimers — Au@Pd and Au@Pt — were synthesized using Au NP trimers as seeds and by reducing Pd or Pt precursors with $NH_2OH·HCl$ at 85 °C for 14 h; based on a modified version of our previously reported procedure (see the SI for details).[46] The TEM and HAADF-STEM-EDS images of the products of the synthesis clearly confirm the formation of core-shell NP trimers (Figure 2). The yield of the synthesis of the Au@Pd and Au@Pt NP trimers were 70 and 74%, respectively (Figure S6a and b in the SI). The Pd and Pt composition in the trimers determined by the EDS were 8 and 5 %, respectively. The average NP size, shell thickness, and gap size were, respectively, 60 ± 6, 1, and 0.57 nm for the Au@Pd NP trimers and 60 ± 6, 1, and 0.58 nm for the Au@Pt NP trimers (Figure S6c-f in the SI). The average gap distance of each Au, Au@Pd, and Au@Pt NP trimers was found to be very similar. This can be due to the role of PVP in the final assembly of constituent NPs, as described previously.[7,8,46] The Au@Pd and Au@Pt NP trimers show the remarkable materials' stability through maintaining their composition and morphology after aging 2 weeks in their reaction solution (Figure S7). On the other hand, the shell thickness of the Au@M NP trimers could be manipulated by controlling the amount of M precursors in the synthesis. For instance, Au@Pd NP trimers with average Pd shell thicknesses of 2 and 5 nm were prepared with of 0.1 mL of 5 and 10 mM $K_2PdCl_4$ (standard protocol = 2.5 mM), respectively (Figure S8, in the SI). As shown in the EDS mapping images of Au@Pd NP trimers with thicker thickness compared to standard protocol one, we can observe distinguishable gaps between constituent Au NPs (Figure S8 in the SI).

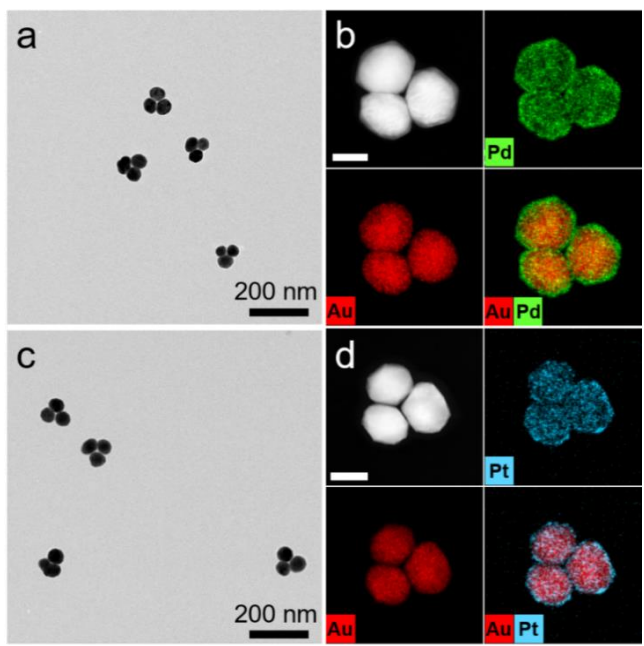

**Figure 2.** TEM images of (a) Au@Pd and (c) Au@Pt NP trimers. HAADF-STEM images and corresponding EDS elemental mapping images of (b) Au@Pd and (d) Au@Pt NP trimers. The scale bars indicate 40 nm.

In order to rationalize the behavior of these structures as plasmonic photocatalysts, we start by studying their optical properties. We measured extinction, absorption, and scattering spectra of samples containing trimers of Au, Au@Pd, and Au@Pt NPs in solution, using a UV-vis extinction/absorption setup coupled to an integrating optical sphere (Figure 3a-c).[51] This combined measurement technique allows for the separation of the scattering and absorption contributions from the standard extinction spectra. The UV-vis extinction spectra of Au, Au@Pd, and Au@Pt NP trimers showed distinct LSPR properties compared to their monomer counterparts (Figure 3a-c and Figure S9 in the SI), which can be attributed to the strong electromagnetic interactions between the NPs, separated by very small gaps.[24,46] For the Au NP trimers, the two main LSPR peaks appeared at 542 and 643 nm. The former can be assigned to the hybridization of the individual dipole plasmon resonance of each constituent NP.[29] As shown in Figure 3a-c, the main LSPR peak, which can be assigned to the Au cores, is damped and blue-shifted when comparing the Au NP trimers (643 nm) to the Au@Pd (625 nm) and

Au@Pt NP (620 nm) trimers due to the presence of Pd and Pt shells.[46,52,53] Here, it is important to note that in all cases (monometallic and hybrid systems) the plasmon modes can be excited, highlighting the ultrathin nature of the catalytic layers (Pd and Pt). Interestingly, we observed that the ratio of absorption over scattering is enhanced from 58% (Au) to 75% (Au@Pd) and 69% (Au@Pt) at the maximum LSPR peak position for each trimer case due to the composition of the shell (Figure 3a-c). This result is not surprising as Pd and Pt are far more lossy metals than Au (the imaginary part of the dielectric function of Pd or Pt is a factor 10–20 higher than that of Au over this spectral range, see Figure S10a in the SI).[54]

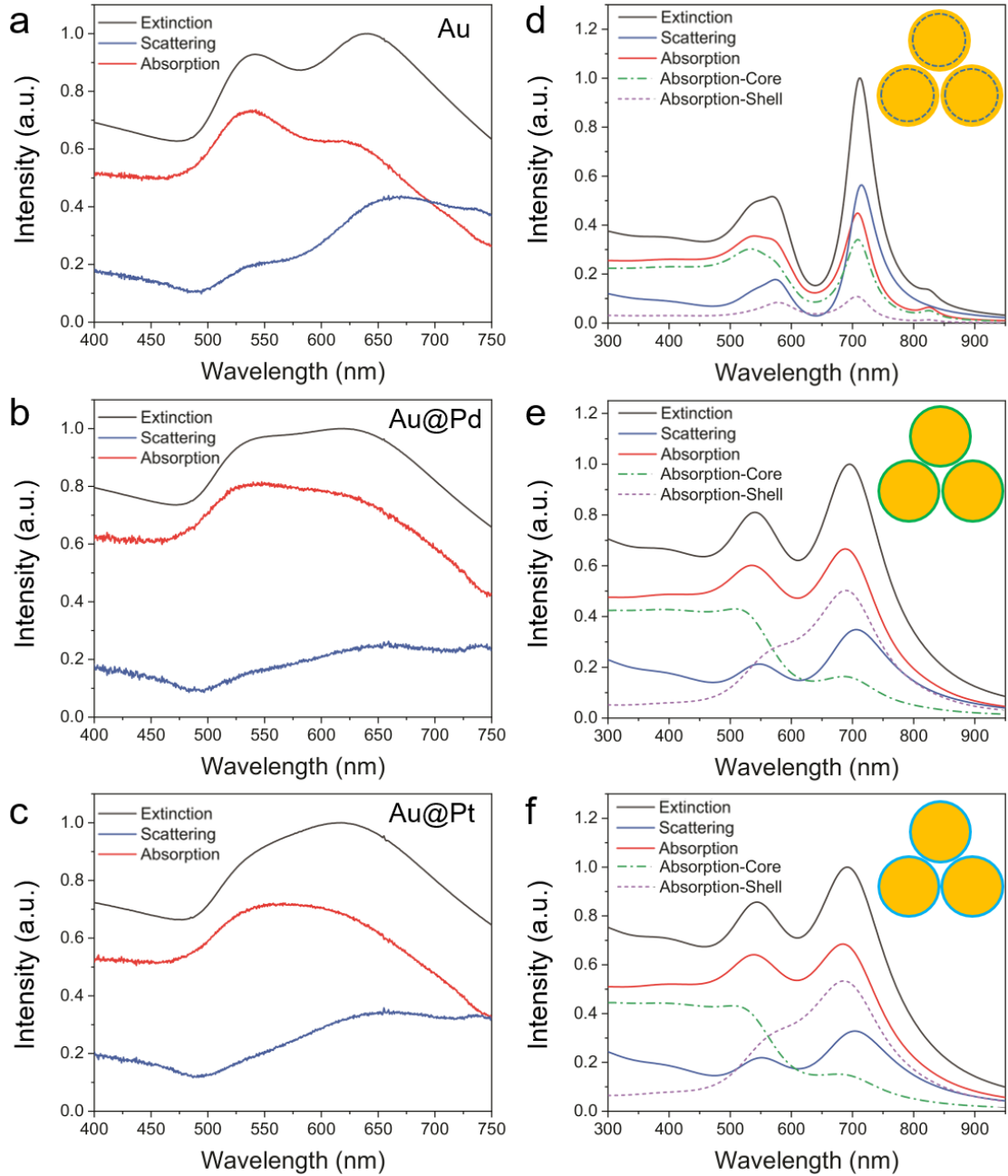

**Figure 3.** Measured fractional extinction, absorption, and scattering spectra of (a) Au, (b) Au@Pd, and (c) Au@Pt NP trimers. Calculated fractional extinction, absorption, and scattering spectra of a model (d) Au, (e) Au@Pd, and (f) Au@Pt NP trimers with orientation-averaged incident light. For ease of comparison of relative scattering and absorption contributions, the data in each plot have been normalized such that extinction peaks at 1.

To interpret further the plasmonic properties of each trimer, we calculated the extinction, absorption, and scattering spectra of Au, Au@Pd, and Au@Pt NP trimers using rigorous electromagnetic simulations. We used the generalized Mie theory combined with the superposition T-matrix method to model the optical response of bimetallic monomers and trimers comprising Au cores and thin shell layers (Pd or Pt).[54–56] A noteworthy strength of this simulation method is the ability to compute orientation-averaged optical properties with minimal extra computational cost. This is an important consideration in the modeling of colloidal samples that are randomly-oriented in solution: as the incident light interacts with the particles with arbitrary direction of incidence and polarization, different resonant modes are excited in the nanostructure and the orientation-averaged response can be different substantially from that obtained under high-symmetry excitation. The results of our simulations are shown in Figure 3d-f (see the SI for further details on the calculation). Each geometric model was built based on the experimentally measured average NP diameter, shell thickness, and gap size (i.e., Au core size: 58 nm, shell thickness: 1 nm, and gap size: 0.6 nm). We observed that the fraction of absorption in the simulated extinction spectra for the Au@Pd and Au@Pt NP trimers is significantly higher than that of the spectra for the Au NP trimers. This is in line with the experimental extinction spectra and absorption/scattering ratio for each NP trimers, as described previously. Further understanding of the origin of the enhanced absorption in the hybrid NP trimers requires decomposing the contribution of the shells and the cores in each case, as follows.

It has been recently shown that core-shell bimetallic NPs allow efficient energy transfer by increasing the overall light absorption and tuning the LSPR decay mechanism.[43–45,47] Linic and co-workers, for example, showed that the ratio of the imaginary part of the dielectric function ($\varepsilon_2$) of the constituent materials, $\varepsilon_2$[core]/$\varepsilon_2$[shell], is a good metric to evaluate the energy flow between plasmonic and catalytic core-shell nanostructures.[43] This is a simple but effective way to account for the probability of the optical excitation of the constituents of a bimetallic structure (i.e., the absorption in a hybrid system). For example, a Ag@Pt nanocube is better in transferring energy to the outmost layer of the NP compared to a pure Ag nanocube (i.e., in this last case, we can imagine a Ag core coated with a thin Ag shell).[43,44] However, when the plasmonic core is changed to Au, the new Au@Pt nanocube is not effective in transferring energy to the Pt shell, in comparison to the Ag@Pt one.[43] This behavior can be related to the ratio of $\varepsilon_2$ for Pt and Au at the main LSPR position for the nanocube, which is relatively low compared to that of Ag and Pt.[43] However, and intriguingly, we obtained that the fraction of absorption occurring within the volume of a 1 nm outer spherical shell of the Au NP trimers increases significantly upon the introduction of both Pd and Pt shells, as shown in the theoretical simulations in Figure 3d in comparison to Figure 3e and f. We estimate a 4.8- and 5.1-fold enhancement in the fraction of shell absorption for the Au@Pd and Au@Pt NP trimers compared to the Au NP trimers at the main LSPR position, respectively. These enhancements can be due to two main factors: the LSPR excitation frequency of the NP trimers placed in the spectral region where $\varepsilon_2$ is high (compared to, for example, the nanocubes) and/or the highly enhanced e-field attained around the inter-particle gaps in the NP trimer structure compared to single particles (Figure S10 in the SI).[43]

To clarify which of the two reasons dominates the enhanced shell absorption, we simulated the optical behavior of a (25 nm SiO$_2$)@(4 nm Au) NP coated with either 1 nm of Au, 1 nm of Pd or 1 nm of Pt. Note that this SiO$_2$@Au NP possesses the LSPR maxima at a wavelength similar to that of the NP trimer structure and with a spherical geometry, but without the hotspots contribution (Figure S11 in the SI). Unlike the NP trimer cases, the total absorption is largely damped after coating the SiO$_2$@Au NP with the Pd or the Pt shell (Figure S11a and b in the SI). This is in line with the previously reported behavior for Au@Pt nanocubes,[43] and we conclude that the spectral position of the resonance alone cannot explain the enhanced absorption in the shell for our bimetallic NP trimers. Furthermore, the enhancement in the shell absorption for the NP trimer is larger than that of SiO$_2$@Au NP after the Pd or the Pt shells are doped into each nanostructure (Figure S11c and d in the SI). From these simulations, we deduce that assembled bimetallic NPs with hotspots enhance the absorption process in the outer shell of the hybrid oligomer. This outer shell enhanced absorption effect could turn central for applications like photocatalysis, as chemical reactions occur at the surface of the NPs.

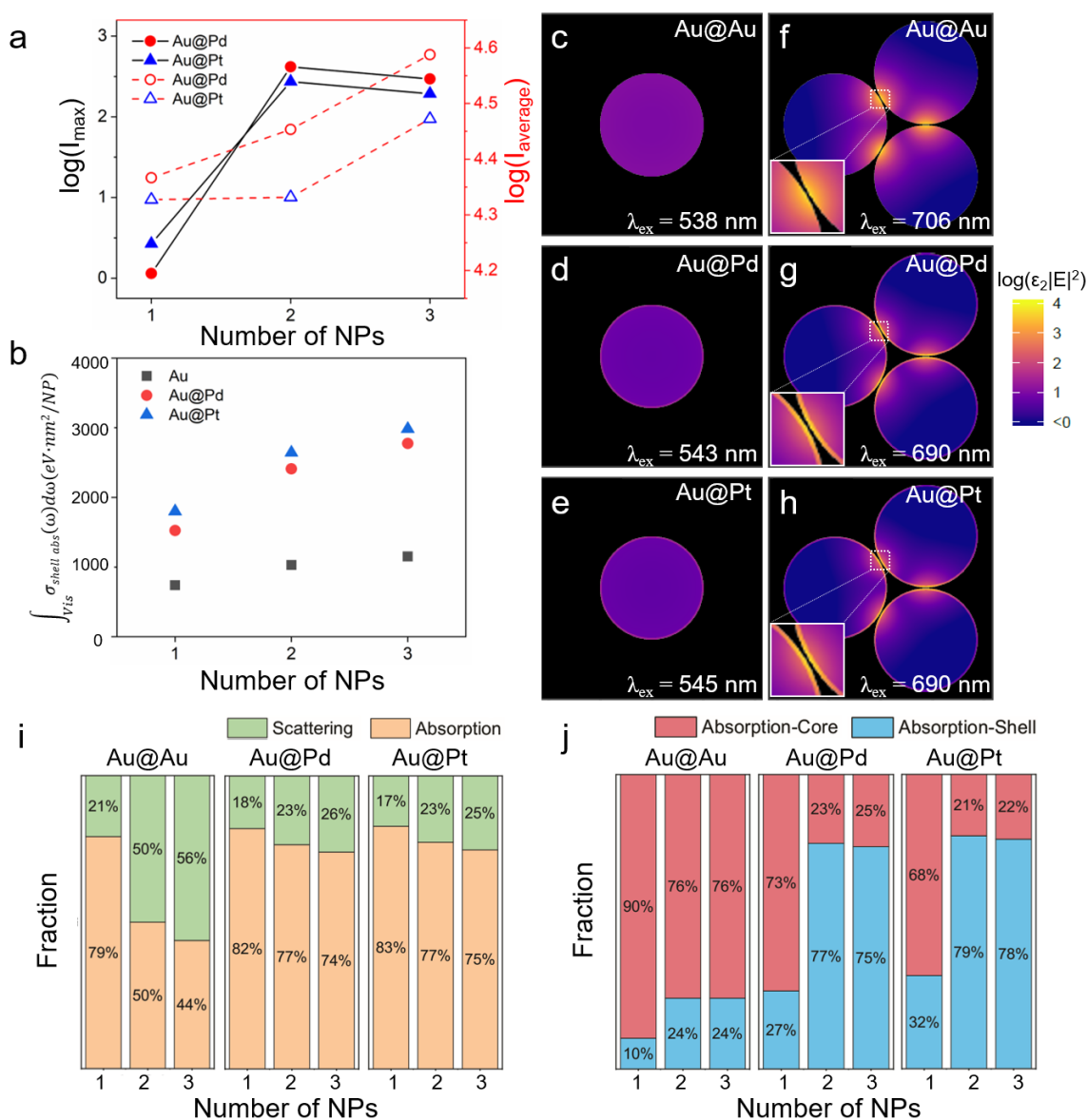

**Figure 4.** (a) Calculated maximum and average of e-field intensity in the outer 1-nm shell at the maximum LSPR peak of each nanostructure. (b) Calculated shell absorption per NP for each nanostructure, integrated over the visible range (from 400 to 750 nm). Calculated absorption mapping images of a model (c) Au, (d) Au@Pd, and (e) Au@Pt NP monomers, and (f) Au, (g) Au@Pd, and (h) Au@Pt NP trimers illuminated with circularly polarized light incident normal to the plane. The spheres are 60 nm in outer diameter. (i) The ratio of absorption and scattering, and (j) the ratio of core absorption and shell absorption in each nanostructure at each of the main LSPR peak position.

To evaluate quantitatively the influence of the NP-assembly on their photochemical reactivity, we investigated next the trend of e-field intensity and associated absorption in the outermost thin-shell for monomers, dimers, and trimers (Figure 4 and Figures S12 in the SI). Note that these parameters are strongly correlated with the observed ability of plasmons to induce chemical transformations.[18,21,24,26,40,43–47] As a comparative measure for the different oligomers, we calculated the average e-field intensity over the entire 1-nm outer shell for a monomer, dimer, or trimer, as shown in Figure 4a, and the spectrally-integrated (from 400 to 750 nm) absorption cross-section of the 1 nm outer-shell for each structure, Figure 4b. These results are extended not only to the Au NP cases but also to the bimetallic ones, such as Au@Pd and Au@Pt.

As shown in Figure 4a and b, the integrated e-field intensity and absorption in the external shell of each nanostructure increase with the number of NPs, as expected. Specifically, the Au@Pd and Au@Pt NP trimers exhibit a 256 and 72-fold enhancement of the maximum e-field intensity compared to their respective NP monomer and a 1.6 and 1.4-fold enhancement of the average e-field intensity (averaged over-illumination angles and outer shell volume), respectively. As shown by Govorov and co-workers, in a nanostructure with a hotspot, 50% of the hot-carriers generation rate can be attributed to the hotspot region, even if it only accounts for a 4% of the total surface area.[57] This is mainly due to the enormous e-field confinement achieved inside the metal nanostructures in the hotspot area. This is proportional to the e-field outside the nanostructures, as shown by the maximum e-field values for our trimer structures (Figure 4a and f-h). Note that the maximum value for the e-field between dimers and trimers is not very different. However, trimers have two hotspots per NP while dimers have only one. This again points towards an enhanced production of carriers for the hybrid trimers compared to their monometallic and monomer/dimer hybrid cases. In the same line, the Au@Pd and Au@Pt NP trimers exhibit 5.4 and 6.3 times increase of the shell-absorption cross-section with respect to their monomeric counterparts. This implies that each bimetallic NP that constitutes the trimer absorbs in its shell 1.8 (Au@Pd) and 2.1 (Au@Pt) times more than the same isolated NP (Figure 4b). This reinforces the importance of the structure and the hotspots for enhancing photon absorption in the catalytic shell of these hybrid NPOs. Our results demonstrate the central role of hotspots in dominating the optical and subsequently the catalytic response in these hybrid oligomeric structures. The Au@Pd and Au@Pt NP trimers also exhibit 2.4 and 2.6 times increase of the shell-absorption with respect to Au NP trimers (Figure 4b); which points as well to the role of composition — besides structure — in hybrid systems like these ones. These enhancements can be due to the LSPR excitation frequency of the bimetallic trimers placed in the spectral region where $\varepsilon_2$ is high compared to Au NP trimers (Figure S10a in the SI), as already described by Chavez and co-workers.[43]

Figure 4c-h show the maps of the near-field intensity distribution for monomers and trimers at their main LSPR maxima (see the SI). The color scale corresponds to $\varepsilon_2$ times the $|E|^2$ — $\log(\varepsilon_2|E|^2)$ — rather than the more customary local field intensity ($|E|^2$), as our objective is to understand the local rate of absorption, which in turn influences the generation of hot electrons.[18,21,24,26] The regions of high absorption are more strongly localized within the Pd or Pt shell for bimetallic trimers than for the pure Au NP trimers, as evidenced in the insets of Figure 4f-h. Additional simulations for dimers reveal that trimers are better able to concentrate the absorption into the shell region from supporting two hotspots per particle, compared to a single hotspot for dimers (Figure S13 in the SI). Figure 4i and j show the ratio of absorption and shell absorption in each nanostructure at their main LSPR peak position. The core-shell bimetallic nanostructures preserve well their portion of absorption in total extinction as increasing the number of NPs compared to the monometallic nanostructures (Figure 4i).

Furthermore, the core-shell bimetallic nanostructures can change the dominant absorption pathway from core to shell absorption as increasing the number of NPs compared to the monometallic ones (Figure 4j). From these figures, we can obviously check that the synergy effect of bimetallic composition and hot spots as inducing an increment of non-radiative plasmon decay to the surface. These simulations suggest that the formation of the bimetallic NP trimers is a favorable strategy for enhancing the absorption of light in the thin catalytic metal shell. This can be achieved by combining the light-harvesting ability of plasmon resonances from the core particle with the localized absorption in the shell; arising from its lossy dielectric function and the multiple hotspots between neighboring particles.[58]

We have so far described the synthesis and optical properties of the hybrid trimers towards their usage as plasmonic photocatalysts. However, the next central aspect of revisiting is their stability in reaction conditions. The stability of colloidal nanostructures has been a critical issue for their application as colloidal catalysts. Most of the reported results in the literature deal with supported NPs to mitigate their weak in-operando stability.[4,15,18,33,34,39,41,42] Similar to sintering in traditional catalysts, for colloidal catalysts the main problem is their fast aggregation under reaction conditions. This process reduces considerably the catalytic activity and the available surface area for colloidal catalysts. As such, stable and well-dispersed colloidal photocatalysts can open new paths for large-scale applications. To assess the stability of our samples, we measured the UV-vis spectra and TEM images of the Au, Au@Pd, and Au@Pt NP trimers in various conditions such as ultrasonic treatment, pH 4, pH 10, and ethanol (EtOH) for a prolonged period of time (Figures S14-17 in the SI). Observing the change of the LSPR peak position of the NPs is a simple and reliable way to verify the stability of nanostructures, as it is very sensitive to minute changes in shape or composition.[1,59] In these tests, we have verified that there is no significant change in the LSPR peak position when changing the reaction conditions for any of the studied trimer systems. This remarkable structural stability can be attributed to the abovementioned strong van der Waals interactions between NPs in the trimers configuration.

Finally, we conclude our study by testing the performance of the synthesized bimetallic Au@Pd NP trimers as photocatalysts. As a model system, we tested the photocatalytic reduction of 4-NBT, which has become a popular test system for studying hot electron-induced chemical conversion, using surface enhanced Raman scattering (SERS) as a detection method.[26,60–65] In order to benchmark our bimetallic NP trimers, we synthesized Au and Au@Pd NP monomers to isolate the effect of the hotspots in photocatalytic response of these systems (see the Experimental details and Figure S18 in the SI). For the SERS measurements, all the samples were centrifuged and washed with deionized water three times to remove the remaining surfactant molecules. Figure 5 and Figure S19 in the SI show the time-resolved SERS spectra of 4-NBT ($10^{-5}$ M) on Au and Au@Pd NP trimers and monomers in aqueous solution, respectively, using a 633 nm laser irradiation at a power of 2 mW $cm^{-2}$. At time zero (t = 0 s), the SERS spectra of each NP trimers and monomers show the three major vibration peaks of 4-NBT at 1084, 1334, and 1569 $cm^{-1}$, assigned to the C-S stretching, $NO_2$ symmetric stretching, and C=C stretching modes, respectively. As the reaction proceeds, new SERS peaks at 1145, 1393, and 1439 $cm^{-1}$ — assigned to the vibrational modes of DMAB — gradually appear in the Au@Pd NP trimers case (Figure 5b).[58–61] Notably, the Au@Pd NP trimers show distinguishable photo-induced reduction of 4-NBT in aqueous solution compared to the Au NP trimers and the Au and Au@Pd monomers for the same reaction conditions and illumination powers (Figure 5c). These results highlight that for these short illumination times and powers, the plasmonic-catalytic NP trimer configuration is the only one capable of driving the reaction. This confirms the synergy between structure and composition as a promising strategy in designing future plasmonic catalysts.

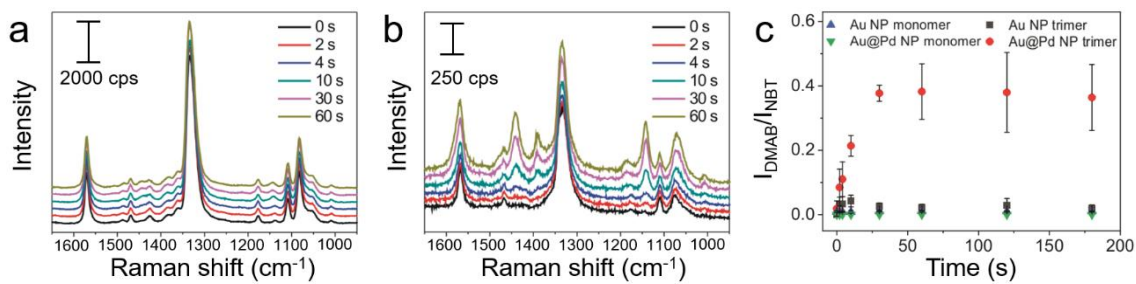

**Figure 5.** Time-dependent SERS spectra of 4-NBT obtained with (a) Au and (b) Au@Pd NP trimers under a 633 nm irradiation in aqueous solution. (c) Time traces of the ratio of 4-NBT (1334 cm$^{-1}$) and DMAB (1145 cm$^{-1}$) peak intensities for each nanostructure.

In summary, we report a novel synthesis strategy for the realization of Au and bimetallic (Au@Pd and Au@Pt) core-shell NP trimers with remarkable structural stability in aqueous solution. The well-defined morphology of the Au NP trimer arises from the sacrificial template assembly strategy by finely controlling the galvanic replacement of Ag nanoprisms with Au precursors at very low temperature. We studied in detail the optical and photocatalytic properties of these new structures and found that the assembly of bimetallic core-shell NPs induces an enhancement in the light absorption within the catalytic metal shells, and consequently assisting the chemical transformation process of molecular adsorbates. According to our experimental studies and theoretical analysis, this strong localized absorption results not only from the bimetallic composition (material and thickness) but also from the presence of electromagnetic hotspots associated with the gaps between constituent NPs. We used SERS measurements to monitor the plasmon-mediated reduction of 4-NBT in aqueous solution, demonstrating the plasmonic-catalytic synergy of core-shell NPOs, which outperform their monomer or monometallic counterparts. Furthermore, the remarkable structural stability of NP trimers in various conditions will give us chance to support over a variety of other matrices for further catalytic studies and enable easy re-use. We envision that the present study can provide a new direction toward the development of efficient photocatalysts by exploring new types of plasmonic-catalytic oligomers.

■ ASSOCIATED CONTENT

**Supporting Information**

Experimental details, SEM images, TEM images, UV-vis spectra, and calculated extinction, absorption, and scattering spectra of samples, UV-vis extinction spectra of NP trimers in various conditions, TEM images and EDS elemental mapping images of NP monomers, and SERS spectra of 4-NBT obtained with NP monomers.

■ AUTHOR INFORMATION

**Corresponding Authors**

**Emiliano Cortés** – *Chair in Hybrid Nanosystems, Nanoinstitute Munich, Faculty of Physics, Ludwig-Maximilians-Universität München, 80539 Munich, Germany*; orcid.org/0000-0001-8248-4165; Email: Emiliano.Cortes@lmu.de

**Sang Woo Han** – *Center for Nanotectonics, Department of Chemistry and KI for the NanoCentury, KAIST, Daejeon 34141, Korea*; orcid.org/0000-0001-8248-4165; Email: sangwoohan@kaist.ac.kr

**Authors**

**Seunghoon Lee** – *Center for Nanotectonics, Department of Chemistry and KI for the NanoCentury, KAIST, Daejeon 34141, Korea; Nanoinstitute Munich, Faculty of Physics, Ludwig-Maximilians-Universität München, 80539 Munich, Germany*; orcid.org/0000-0001-9447-4503

**Heeyeon Hwang** – *Center for Nanotectonics, Department of Chemistry and KI for the NanoCentury, KAIST, Daejeon 34141, Korea*

**Wonseok Lee** – *Center for Nanotectonics, Department of Chemistry and KI for the NanoCentury, KAIST, Daejeon 34141, Korea*

**Dmitri Scherbarchov** – *The MacDiarmid Institute for Advanced Materials and Nanotechnology, School of Chemical and Physical Sciences, Victoria University of Wellington, P.O. Box 600, Wellington 6140, New Zealand*; orcid.org/0000-0002-8385-7186

**Younghyun Wy** – *Center for Nanotectonics, Department of Chemistry and KI for the NanoCentury, KAIST, Daejeon 34141, Korea*

**Johan Grand** – *The MacDiarmid Institute for Advanced Materials and Nanotechnology, School of Chemical and Physical Sciences, Victoria University of Wellington, P.O. Box 600, Wellington 6140, New Zealand; Université de Paris, ITODYS, CNRS, F-75006, Paris, France*; orcid.org/0000-0001-6920-5184

**Baptiste Auguié** – *The MacDiarmid Institute for Advanced Materials and Nanotechnology, School of Chemical and Physical Sciences, Victoria University of Wellington, P.O. Box 600, Wellington 6140, New Zealand*; orcid.org/0000-0002-2749-5715

**Dae Han Wi** – *Center for Nanotectonics, Department of Chemistry and KI for the NanoCentury, KAIST, Daejeon 34141, Korea*

**Notes**

The authors declare no competing financial interest.

■ **ACKNOWLEDGMENTS**

This work was supported by the National Research Foundation of Korea (NRF) grant funded by the Korea government (MSIT) (2015R1A3A2033469, 2018R1A5A1025208). E.C. acknowledges funding and support from the Deutsche Forschungsgemeinschaft (DFG, German Research Foundation) under Germany´s Excellence Strategy – EXC 2089/1 – 390776260, the Bavarian program Solar Energies Go Hybrid (SolTech), the Center for NanoScience (CeNS) and the European Commission through the ERC Starting Grant CATALIGHT (802989).

*Res.* **2019**, *52*, 2784−2792.